%% file: etah_summer_2005.tex
\documentclass[aps,prl,preprint,tightenlines,superscriptaddress,showpacs,byrevtex]{revtex4}

\usepackage{graphicx} % Include figure files
\usepackage{dcolumn}  % Align table columns on decimal point
\usepackage{xspace}
\usepackage{color}
\graphicspath{{ps}}

\newcommand{\bb}{\ensuremath{B \overline{B}}\xspace}

\newcommand{\btoetapi}{\ensuremath{B^\pm \to \eta \pi^\pm}\xspace}
\newcommand{\btoetak}{\ensuremath{B^\pm \to \eta K^\pm}\xspace}

\newcommand{\btoetakz}{\ensuremath{B^0 \to \eta K^0}\xspace}

\newcommand{\etagg}{\ensuremath{\eta_{\gamma \gamma}}\xspace}
\newcommand{\etapi}{\ensuremath{\eta_{3\pi}}\xspace}

\newcommand{\de}{\ensuremath{\Delta E}\xspace}
\newcommand{\br}{\ensuremath{\mathcal{B}}\xspace}
\newcommand{\acp}{\ensuremath{\mathcal{A}_{CP}}\xspace}

\def\calL{{\mathcal L}}

\def\Mbc{M_{\rm bc}}
\def\LR{{\mathcal R}}

\begin{document}

%\vspace*{-3\baselineskip}
%\resizebox{!}{3cm}{\includegraphics{belle.eps}}

\preprint{\vbox{ %\hbox{   }
                 \hbox{BELLE-CONF-0525}
                 \hbox{EPS05 497} 
                 %\hbox{hep-ex nnnn, if available}
}}

\title{ \quad\\[0.5cm] Improved Measurements of Branching Fractions and 
$CP$ Asymmetries in $B\to \eta h$ Decays}

%%%% >>>>> insert the authorlist here. BEFORE the abstract !!!!! <<<<<
\input{author-conf2005.tex}
%\collaboration{Belle Collaboration}
%\noaffiliation
%\collaboration{The Belle Collaboration}

\begin{abstract}
We report improved measurements of $B$ decays 
with an $\eta$ meson in the final state using $357 \mathrm{~fb}^{-1}$ of 
data collected by the Belle detector at the KEKB $e^+ e^-$ collider.  
We observe the decays  $\btoetapi$ and   $\btoetak$; the 
measured branching fractions are
$\br(\btoetapi) = ( 3.9\pm 0.5 \mathrm{(stat)} \pm 0.2
 \mathrm{(sys)}) \times 10^{-6}$ and
$\br(\btoetak) = ( 2.2\pm 0.4 \mathrm{(stat)} \pm 0.1
\mathrm{(sys)}) \times 10^{-6}$. Their corresponding $CP$-violating
asymmetries are measured to be $-0.10\pm 0.11 \mathrm{(stat)} 
\pm 0.02 \mathrm{(sys)}$ for $\eta \pi^\pm$ and $-0.55\pm 0.19 \mathrm{(stat)} 
^{+0.04}_{-0.03} \mathrm{(sys)}$ for $\eta K^\pm$. No significant signal 
is found for $B^0\to \eta K^0$ decays; the upper limit on the 
branching fraction at the 90\% confidence level is
$\br(\btoetakz)< 1.9\times 10^{-6}$.

\end{abstract}


\maketitle

%%%% >>>> keep the final version single-spaced
\tighten

{\renewcommand{\thefootnote}{\fnsymbol{footnote}}}
\setcounter{footnote}{0}
Charmless $B$ decays provide a  rich sample to understand  $B$ decay
dynamics and to search for $CP$ violation. It has been suggested \cite{theory}
that direct $CP$ violation could be large in $B^+\to\eta \pi^+$ \cite{CC} and 
$B^+\to\eta K^+$ decays due to penguin-tree  interference. In our earlier
measurements \cite{etahprd}, 
no $CP$-violating  asymmetry ($A_{CP}$) was seen for the 
$\eta \pi^+$ mode.  Belle's central value of $A_{CP}$  for
$\eta K^+$ was large in magnitude and negative, but statistics were 
limited even for
the branching fraction measurement. In the most recent update from the BaBar 
collaboration \cite{babar}, the previous negative $A_{CP}$ values with
$\sim 2 \sigma$ significance for both $\eta K^+$ and $\eta \pi^+$ modes
 have moved
to within $\sim 1 \sigma$ deviation from zero.  Because experimental 
uncertainties
on $CP$ asymmetries and branching fractions  are still large, it is 
necessary to analyze larger data sample.     

In this paper, we report improved  measurements of branching fractions and 
partial rate asymmetries for  $B\to\eta h$ decays, where $h$ is a 
$K$ or $\pi$  meson. The partial rate asymmetry for
charged $B$ decays is  defined to be:
\begin{eqnarray}
\acp=\frac{N( B^- \to \eta h^-)-N(B^+ \to \eta h^+)}
{N(B^- \to \eta h^-)+N(B^+ \to \eta h^+)},
\end{eqnarray}
where $N(B^-)$ is the yield for the $B^- \to \eta h^-$ decay and
$N(B^+)$ denotes that of the charge conjugate mode.
 The data sample consists of 386
million \bb pairs (357 fb$^{-1}$) collected
with the Belle detector at the KEKB $e^+e^-$ asymmetric-energy  
(3.5 on 8~GeV) collider~\cite{KEKB} operating at the $\Upsilon(4S)$ resonance.

The Belle detector is a large-solid-angle magnetic
spectrometer that
consists of a silicon vertex detector (SVD),
a 50-layer central drift chamber (CDC), an array of
aerogel threshold \v{C}erenkov counters (ACC),
a barrel-like arrangement of time-of-flight
scintillation counters (TOF), and an electromagnetic calorimeter (ECL)
comprised of CsI(Tl) crystals located inside
a superconducting solenoid coil that provides a 1.5~T
magnetic field.  An iron flux-return located outside of
the coil is instrumented to detect $K_L^0$ mesons and to identify
muons (KLM).  The detector
is described in detail elsewhere~\cite{Belle}.
In August 2003, the three-layer SVD was replaced by a four-layer radiation
tolerant device. The data sample used in this analysis consists of 
140 fb$^{-1}$ of data with the old SVD (Set I) and 
217 fb$^{-1}$ with the new one (Set II).

The event selection and candidate reconstruction are similar to 
that documented in our previous paper \cite{etahprd}. Here, we only give 
a brief description of event selection. 
Two $\eta$ decay channels are considered in this analysis: 
$\eta\to \gamma\gamma$ ($\eta_{\gamma\gamma}$) and $\eta\to \pi^+\pi^-\pi^0$
 ($\etapi$). We require photons from $\eta$ and $\pi^0$ candidates to have 
laboratory energies above 50 MeV. In the $\etagg$ reconstruction, the 
energy asymmetry,
defined as the absolute value of the energy difference in the laboratory frame
between the two photons divided by their energy sum,
must be less than 0.9. Neither photon is allowed
to pair with any other photon with at least 100 MeV energy to form a 
$\pi^0$
candidate. Candidate $\pi^0$ mesons are selected by requiring the
two-photon invariant mass to be in the mass window between 115 MeV/$c^2$ and
152 MeV/$c^2$.  The momentum vector of each photon is then readjusted to
constrain the mass of the photon pair to the nominal $\pi^0$ mass.

Candidate $\etapi$ mesons are reconstructed by combining a $\pi^0$ with at
least 250 MeV/$c$ laboratory momentum with a pair of oppositely charged 
tracks that 
originate from the interaction point (IP). 
We make the following requirements on the invariant mass of 
the $\eta$ candidates in both data sets: 
516 MeV/$c^2 < M_{\gamma\gamma} < 569$ MeV/$c^2$ for 
$\etagg$ and 539 MeV/$c^2 < M_{3\pi} <556$ MeV/$c^2$ for $\etapi$. After the 
selection of each candidate, the $\eta$ mass constraint is implemented  
by readjusting the momentum vectors 
of the daughter particles.   

Charged tracks are required to come from the IP.  Charged kaons and pions,
which are combined with $\eta$ mesons to form $B$ candidates,
are identified using a $K(\pi)$  likelihood $L_K(L_\pi)$ obtained 
by combining information from the CDC ($dE/dx$),
the TOF and the ACC. Discrimination between kaons and pions is achieved 
through a  requirement on the likelihood ratio $L_{K}$/($L_{\pi}+L_{K}$). 
Charged tracks with  
likelihood ratios greater than 0.6 are regarded as kaons, and less than 0.4
as pions. Furthermore, charged tracks that are positively identified as 
electrons or muons are rejected. The $K/\pi$ identification efficiencies 
and misidentification rates are determined from a sample of 
$D^{*+}\to D^0\pi^+, D^0\to K^-\pi^+$ decays. In reference \cite{acpkpi}
we reported that in both data sets  $\pi^+$ has a higher efficiency than 
$\pi^-$ while the
kaon identification efficiency is higher for $K^-$ than $K^+$. The efficiency
difference introduces a bias in $A_{CP}$ which needs to be corrected.     
$K^0_S$ candidates 
are reconstructed from  pairs of oppositely-charged tracks
with an invariant mass ($M_{\pi\pi}$) between 480 and 516 MeV/$c^2$.
Each candidate must have a displaced vertex with a flight direction
consistent with that of a $K^0_S$ originating from the IP.

Candidate $B$ mesons are identified using the beam constrained mass,
$\Mbc =  \sqrt{E^2_{\mbox{\scriptsize beam}} - P_B^2}$,
and the energy difference, $\Delta E = E_B  - E_{\mbox{\scriptsize beam}}$,
where $E_{\mbox{\scriptsize beam}}$ is the run-dependent beam energy in the
$\Upsilon(4S)$ rest frame and  is determined  from
$B\to D^{(*)}\pi$ events,
and $P_{B}$ and $E_B$ are the momentum and energy of  the
$B$ candidate in the $\Upsilon(4S)$ rest frame. The resolutions on $\Mbc$ and 
$\de$ are about 3 MeV/$c^2$ and 20--30 MeV, respectively.  
Events with $\Mbc >5.2$ GeV/$c^2$ and $|\de|<0.3$ GeV are selected for the 
 analysis.   

The dominant background comes from the $e^+e^-\rightarrow q\bar{q}$ continuum, 
where $q= u, d, s$ or $c$. To distinguish signal from the jet-like continuum 
background,  event shape variables and $B$ flavor tagging information 
are employed. We combine information of correlated shape variables into a 
Fisher discriminant \cite{fisher} and compute the likelihood as a product of
probabilities of this discriminant and $\cos\theta_B$, where $\theta_B$ is 
the angle between the $B$ flight direction and the beam direction
in the $\Upsilon(4S)$ rest frame. A likelihood ratio, 
$\LR = {\calL}_s/({\calL}_s + {\calL}_{q \bar{q}})$, is formed from signal
(${\calL}_s$) and background (${\calL}_{q \bar{q}}$) likelihoods, obtained 
using events from the signal Monte Carlo (MC) and from data with 
$\Mbc< 5.26$ GeV/$c^2$, respectively.  Additional 
background discrimination is provided by $B$ flavor tagging. An event that 
contains a lepton (high quality tagging) is more likely to be a 
$B \overline B$ event so a looser $\LR$ requirement can be applied. 
We divide the data into six sub-samples based on 
the quality of flavor tagging \cite{tagging} (see details in \cite{etahprd}).
Continuum suppression is achieved by applying a mode dependent requirement on 
$\LR$ for events in each sub-sample in Set I and Set II according to  
$N_s^{\rm exp}/\sqrt{N_s^{\rm exp}+N_{q\bar{q}}^{\rm exp}}$,
where $N_s^{\rm exp}$ is the expected signal  from MC and
$N_{q\bar{q}}^{\rm exp}$ denotes the number of background events estimated from data.

From MC all other backgrounds are found to be negligible 
 except for the $\eta K^+ \leftrightarrow \eta \pi^+$ reflection, due to 
$K^+\leftrightarrow \pi^+$ 
misidentification, and the feed-down from charmless $B$ decays, predominantly  
$B\to \eta K^*(892)$ and  $B\to \eta\rho(770)$. 
We include these two components in the fit used 
to extract the signal.   

The signal yields and partial rate asymmetries
are obtained using an extended 
unbinned maximum-likelihood (ML) fit with input variables $\Mbc$ and $\de$.
The likelihood is defined as: 
\begin{eqnarray} 
\mathcal{L} & = &\rm{exp}^{-\sum_j N_j}
\times \prod_i (\sum_j N_j \mathcal{P}_j) \;\;\; \mbox{and} \\
\mathcal{P}_j & = &\frac{1}{2}[1- q_i \cdot \acp{}_j ]
P_j(M_{{\rm bc}i}, \Delta E_i),
\end{eqnarray}
where $i$ is the identifier of
the $i$-th event, $P(M_{\rm bc}, \Delta E)$ is the two-dimensional probability
density function (PDF) in
$M_{\rm bc}$ and $\Delta E$, $q$ indicates the $B$ meson flavor,
$B^+(q=+1)$ or $B^- (q=-1)$, $N_j$ is the number of events for the
category $j$, which corresponds to either signal, $q\bar{q}$ continuum,
a reflection due to $K$-$\pi$ misidentification, or
background from other charmless  $B$ decays. For the neutral $B$ mode, 
$\mathcal{P}_j$ in the equation above is simply 
$P_j(M_{{\rm bc}i}, \Delta E_i)$ and there is no reflection component.   

Since the efficiency of particle identification is slightly different for 
positively  
and negatively charged particles, the raw asymmetry defined in Eq. 1 
must be corrected. This efficiency 
difference results in an $A_{CP}$ bias of  $-0.005$ (0.005) for $\eta \pi$ 
($\eta K$). The bias is subtracted from the value
obtained for the raw asymmetry.

 The signal peak positions and resolutions in $\Mbc$ and $\de$
are adjusted according to the data-MC differences using large control samples 
of $B\to D\pi$  and $\overline{D}{}^0\to K^+\pi^-\pi^0/\pi^0\pi^0$ decays. 
The continuum background in $\de$ is described by a first or second order
polynomial while the $\Mbc$ distribution is parameterized by an 
ARGUS function, $f(x) = x \sqrt{1-x^2}\;{\rm exp}\;[ -\xi (1-x^2)]$, where 
$x$ is $\Mbc$ divided by half of the total center of mass energy \cite{argus}. 
Thus the continuum PDF is the product of an ARGUS function and a polynomial, 
where $\xi$ and the coefficients of the polynomial are free parameters. 
The PDFs of the reflection and charmless $B$ backgrounds are modelled as 
two-dimensional smooth functions, obtained using large MC samples.   
 
All the signal and background yields and their partial rate asymmetries 
are free parameters in the fit except for the reflection components, where
the $A_{CP}$ and the normalizations  are fixed to expectations 
based on the $B^+\to \eta K^+$ and $B^+\to \eta \pi^+$  partial rate 
asymmetries and branching fractions, as well as   
$K^+\leftrightarrow \pi^+$ fake rates. The reflection yield and $A_{CP}$ 
are first input with the assumed values and are then
recalculated according to our measured results.

\begin{table*}[th]
\caption{ Detection efficiency ($\epsilon$) including sub-decay branching 
fraction, yield,
significance (Sig.), measured branching fraction ($\mathcal{B}$), 
the 90\% C.L. upper limit (UL) and $A_{CP}$ for the $B\to \eta h$ decays. 
The first errors
in columns 3, 5 and 7 are statistical and the second errors are systematic.}

\begin{tabular}{lcccccc} \hline\hline
Mode & $\epsilon(\%)$ &    Yield & \hspace{0.2cm}Sig. 
\hspace{0.2cm} & $\mathcal{B} (10^{-6})$  & UL$(10^{-6}$)& $A_{CP}$ \\
\hline
\hline
 \btoetapi &
        & & 10.9& $3.9\pm 0.5\pm 0.2$ &
         & $-0.10\pm 0.11 \pm 0.02$
        \\
 \hspace{0.2cm} $\etagg \pi^\pm$  & & & 8.7 &$3.9^{+0.6}_{-0.5}\pm 0.2$ & 
                &$0.08\pm 0.14 \pm 0.02$ \\
\hspace{0.6cm} Set I&
       $9.12$  & $71.5^{+13.3}_{-12.5}\pm 2.0$ & 7.0 & $5.2^{+1.0}_{-0.9}
\pm 0.3$  &  & $ 0.14 \pm 0.18\pm 0.02$
        \\
\hspace{0.6cm} Set II&
       $9.34$ & $68.3^{+15.8+2.7}_{-14.9-2.9}$ & 5.1 & $3.1 \pm 0.7 \pm 0.2$ 
         &  & $ 0.00 \pm 0.22 ^{+0.02}_{-0.03}$  \\    
\hspace{0.2cm}$\eta_{3\pi} \pi^\pm$ & & & 6.7 & $3.8^{+0.9}_{-0.8} \pm 0.3$ & & 
$-0.54\pm 0.22 \pm 0.02$ \\
\hspace{0.6cm}Set I &
         3.26& $16.1^{+6.6}_{-5.7}\pm 0.8$ & 3.3 & $3.2^{+1.3}_{-1.2}\pm 0.3$
  &          & $-0.10\pm 0.38^{+0.02}_{-0.03}$
        \\
\hspace{0.6cm}Set II &
         3.53 &$33.7^{+9.3+2.0}_{5.8-2.2}$ &5.8 & $4.1^{+1.1}_{-1.0}\pm 0.3$
      & & $-0.76^{+0.26+0.03}_{-0.27-0.04}$ \\ 
 \btoetak &
      &  & 7.1 & $2.2\pm 0.4 \pm 0.1$ &  & $-0.55\pm 0.19^{+0.04}_{-0.03}$ \\
\hspace{0.2cm}$\etagg K^\pm$ &
             & & $5.8$ & $2.0\pm 0.5^{+0.2}_{-0.1}$& & $-0.71^{+0.24+0.05}_{-0.27-0.04}$  \\
  \hspace{0.6cm} Set I &
         $8.35$ & $27.0^{+10.0+2.1}_{-9.0-2.3}$ & $3.3$ & $2.1^{+0.8}_{-0.7}\pm
       0.2$ &  & $-0.44^{+0.34}_{-0.37}\pm 0.03$ \\
  \hspace{0.6cm} Set II & 
         $8.47$ & $39.0^{+13.0+2.7}_{-12.1-2.8}$ & 4.5 & 
    $2.0^{+0.7}_{-0.6}\pm 0.2 $ & & $-0.94^{+0.32}_{-0.41}\pm 0.07$ \\
\hspace{0.2cm}$\eta_{3\pi} K^\pm$  &  
        & & 4.1 & $2.3^{+0.8+0.2}_{-0.6-0.1}$ & & $-0.28^{+0.29+0.04}_{-0.31-0.03}$ \\         
  \hspace{0.6cm}Set I  
        & $3.03$ & $7.3^{+5.4+1.0}_{-4.3-0.8}$ & 2.2 & 
$1.6^{+1.2}_{-0.9}\pm 0.2$ &  & $-0.66^{+0.67+0.13}_{-1.16-0.10}$ \\
   \hspace{0.6cm}Set II 
        & $3.19$  & $21.2^{+7.7+0.9}_{-6.8-0.8}$ & 3.5 & 
    $2.8^{+1.0}_{-0.9}\pm 0.2$ & & $-0.20^{+0.34}_{-0.33} \pm 0.03$ \\
 $B^0\to \eta K^0$ &
       & & 2.1 & $0.9\pm 0.6\pm 0.1$ & $<1.9$ & \\
\hspace{0.2cm}$\etagg K^0$ & & & 1.0 & $0.6^{+0.7}_{-0.6}\pm 0.1$ &$<2.2$ & \\
  \hspace{0.6cm}Set I  & $3.06$ & $-1.6^{+4.3+0.5}_{-3.1-0.4}$ 
      &$-$  & $-0.3^{+0.9}_{-0.7}\pm 0.1$ &$<1.7$ & \\
   \hspace{0.6cm}Set II & $3.17$ & $9.0^{+7.2}_{-6.3}\pm 0.7$ & 
     1.4 & $1.2^{+1.0}_{-0.8}\pm 0.1$ &$<3.1$ & \\
\hspace{0.2cm}$\eta_{3\pi} K^0$ & & & 1.8 & $1.5^{+1.2}_{-1.0}\pm 0.1$ &$<3.9$  & \\
  \hspace{0.6cm}Set I & $0.90$ & $3.5^{+3.5+0.3}_{-2.6-0.5}$
      &  1.5& $2.3^{+2.5}_{-1.9}\pm  0.3$ &$<6.0$ &\\
  \hspace{0.6cm}Set II & $1.00$ & $2.4^{+3.4+0.2}_{-2.4-0.3}$
      &  1.0& $1.0^{+1.5}_{-1.0} \pm 0.1$ &$<3.7$ &\\
\hline
\hline
\end{tabular}
\label{tab:result}
\end{table*}

Table ~\ref{tab:result} shows the measured branching
 fractions for each decay mode
as well as  other quantities associated with the measurements.
The efficiency for each mode is determined using  MC simulation and
corrected for the discrepancy between data and MC using the control samples. 
Other than the  particle identification performance discrepancy  
reported earlier \cite{etahprd}, our MC slightly overestimates the detection
efficiency of soft $\pi^0$s in Set II, which results in a 4.3\% correction
for the $\etapi$ mode. The combined branching fraction of two data sets
are computed as the sum of the yield divided by its efficiency in each set
divided by the number of $B$ mesons.  
The combined branching fraction of the two $\eta$ decay modes is obtained
from the weighted average assuming the errors are Gaussian. 
Systematic uncertainties in the fit due to the uncertainties in the signal PDFs
are estimated by performing the fit after varying their peak positions and 
resolutions by one standard deviation. In $B^\pm \to \eta \pi^\pm$,
the reflection yields in Set I (Set II) are estimated to be 3.4 (5.5) events 
for the $\etagg$ mode and 1.1 (2.0) for $\etapi$. And in 
$B^\pm \to \eta K^\pm$, the reflection yields are 3.2 (6.5) for $\etagg$ and
1.3 (2.2) for $\etapi$.  The reflection yields and their $A_{CP}$ values 
 are varied by one standard deviation in the fit
to obtain the corresponding systematic uncertainties.
The quadratic 
sum of the deviations from the central value gives the systematic 
uncertainty in the fit,
which ranges from 3\% to 6\%. For each systematic check, the 
statistical significance is taken as the square root of
the difference between the value of $-2\ln\calL$ for zero signal yield and the
best-fit value. We regard the smallest value as our significance including 
the systematic uncertainty. 
The number of $B^+B^-$ and $B^0\overline{B}{}^0$ pairs are assumed to be
equal.

The performance of the $\LR$ requirement is 
studied by 
checking the data-MC efficiency ratio using the $B^+\to \overline{D}{}^0 \pi^+$ 
control sample. The obtained systematic error is 2.0--3.3\%. The
 systematic errors on the charged track reconstruction
are estimated to be around $1$\% per track using  partially
reconstructed $D^*$ events.
 The $\pi^0$ and $\eta_{\gamma\gamma}$ 
reconstruction efficiency is verified by comparing the $\pi^0$
decay angular distribution with the MC prediction, and by measuring the ratio 
of the branching fractions for the two $D$ decay channels: 
$\overline{D}{}^0 \to K^+\pi^-$
and $\overline{D}{}^0 \to K^+\pi^-\pi^0$. We assign a 3.5\% (4.0 \%)
error for the $\pi^0$ and $\etagg$ reconstruction in Set I (Set II). 
The $K_S^0$ reconstruction is verified by 
comparing the  ratio of $D^+\to K_S^0\pi^+$ and $D^+\to K^-\pi^+\pi^+$ yields. 
The resulting $K_S^0$ detection systematic errors for Set I and Set II are 
4.4\% and 4.0\%, respectively. The 
uncertainty in the number of $\bb$ events is 1\%. The final systematic error is
obtained by first summing all correlated errors linearly and then quadratically
summing the uncorrelated errors.

\begin{figure}[htb]
\includegraphics[width=0.8\textwidth]{fig1.epsi}
\caption{$\Mbc$ and $\de$ projections for (a,b) $\btoetapi$, (c,d) $\btoetak$,
and (e,f) $\btoetakz$ decays with 
the $\etagg$ and $\etapi$ modes combined. Open
histograms are data, solid curves are the fit functions, dashed lines show
the continuum contributions and shaded histograms are the feed-down component 
from charmless $B$ decays.  
The small contributions around $\Mbc = 5.28$ GeV/$c^2$ and 
$\de =\pm 0.05$
GeV in (a)-(d) are the reflection backgrounds from $\btoetak$ and $\btoetapi$.  
}
\label{fig:mbde}
\end{figure}

 Figure \ref{fig:mbde} shows the $\Mbc$ and $\de$ projections after requiring
events to satisfy $-0.10$ GeV $<\de<0.08$ GeV and $\Mbc>5.27$ GeV/$c^2$, 
respectively.
No significant signal is observed for the $B^0\to \eta K^0$ meson mode.
Significant signals are observed for charged $B$ decays; the corresponding 
$\Mbc$ and $\de$ projections for $B^+$ and $B^-$ samples are shown in  
 Figs. ~\ref{fig:acppi} and  ~\ref{fig:acpk}. 
The $A_{CP}$ results for the two $\eta$ decay modes in 
two data sets are combined assuming that the errors are Gaussian. No 
significant asymmetry is  observed for the $\eta \pi^\pm$ mode while an excess
 of $B^+$ over $B^-$ decays is seen  in the $\eta K^\pm$ mode,
giving a large negative $A_{CP}$ value.   
   
In summary, we have  observed  $\btoetapi$ and  
$\btoetak$ decays; their branching fractions are measured to be 
$(3.9\pm 0.5 \pm 0.2)\times 10^{-6}$ and $(2.2\pm  0.4 \pm 0.1) \times 
10 ^{-6}$, respectively. 
The central values of our measurements in both modes are 
$1.4 \sigma$ smaller than the results \cite{babar} 
 reported by the BaBar collaboration. Our $A_{CP}$ measurement for  $\btoetapi$ 
,$ -0.10\pm 0.11 \pm 0.02$, is consistent with no asymmetry. However,
the partial rate asymmetry for $\btoetak$
remains large in magnitude and negative, 
$A_{CP} = -0.55\pm 0.19^{+0.04}_{-0.03}$, which is 
$2.9 \sigma$ from zero. 
Larger data  samples are  needed to verify this large $CP$ asymmetry. 
Finally, no significant 
$B^0\to \eta K^0$ signal is found and we assign an upper limit at    
the 90\% confidence level of $\mathcal{B}(B^0\to \eta K^0) < 1.9 \times 10^{-6}$.    
 
\begin{figure}[htb]
\includegraphics[width=0.4\textwidth]{fig2_1.epsi}
\includegraphics[width=0.4\textwidth]{fig2_2.epsi}
\includegraphics[width=0.4\textwidth]{fig2_3.epsi}
\includegraphics[width=0.4\textwidth]{fig2_4.epsi}

\caption{$\Mbc$ and $\de$ projections for (left) $B^-\to\eta \pi^-$ and  
(right) $B^+ \to \eta\pi^+$ with 
the $\etagg$ and $\etapi$ modes combined. Open
histograms are data, solid curves are the fit functions, dashed lines show
the continuum contributions and shaded histograms are the contributions
from charmless $B$ decays.  The small contributions near $\Mbc = 5.28$ GeV/$c^2$ and 
$\de = -0.05$ GeV are the backgrounds from misidentified $\btoetak$ 
(reflections).}  
\label{fig:acppi}
\end{figure}

\begin{figure}[htb]
\includegraphics[width=0.4\textwidth]{fig3_1.epsi}
\includegraphics[width=0.4\textwidth]{fig3_2.epsi}
\includegraphics[width=0.4\textwidth]{fig3_3.epsi}
\includegraphics[width=0.4\textwidth]{fig3_4.epsi}

\caption{$\Mbc$ and $\de$ projections for (left) $B^-\to\eta K^-$ and  
(right) $B^+ \to \eta K^+$ with 
the $\etagg$ and $\etapi$ modes combined. Open
histograms are data, solid curves are the fit functions, dashed lines show
the continuum contributions and shaded histograms are the contributions
from charmless $B$ decays.  The small contributions near $\Mbc = 5.28$ GeV/$c^2$ and 
$\de = 0.05$ GeV are the backgrounds from misidentified $\btoetapi$ 
(reflections).}  
\label{fig:acpk}
\end{figure}
%\section*{Acknowledgments}
%***** Acknowledgments *****
% Please paste this acknowledgement into your latex file. 
 %----------- Long version, for most papers ----------- 
We thank the KEKB group for the excellent operation of the
accelerator, the KEK Cryogenics group for the efficient
operation of the solenoid, and the KEK computer group and
the National Institute of Informatics for valuable computing
and Super-SINET network support. We acknowledge support from
the Ministry of Education, Culture, Sports, Science, and
Technology of Japan and the Japan Society for the Promotion
of Science; the Australian Research Council and the
Australian Department of Education, Science and Training;
the National Science Foundation of China under contract
No.~10175071; the Department of Science and Technology of
India; the BK21 program of the Ministry of Education of
Korea and the CHEP SRC program of the Korea Science and
Engineering Foundation; the Polish State Committee for
Scientific Research under contract No.~2P03B 01324; the
Ministry of Science and Technology of the Russian
Federation; the Ministry of Education, Science and Sport of
the Republic of Slovenia; the National Science Council and
the Ministry of Education of Taiwan; and the U.S.\
Department of Energy.

%-------- Short version, if necessary, for PRL -----------
% currently commented out
%We thank the KEKB group for the excellent operation of the
%accelerator, the KEK Cryogenics group for the efficient
%operation of the solenoid, and the KEK computer group and
%the NII for valuable computing and Super-SINET network
%support.  We acknowledge support from MEXT and JSPS (Japan);
%ARC and DEST (Australia); NSFC (contract No.~10175071,
%China); DST (India); the BK21 program of MOEHRD and the CHEP
%SRC program of KOSEF (Korea); KBN (contract No.~2P03B 01324,
%Poland); MIST (Russia); MESS (Slovenia); NSC and MOE
%(Taiwan); and DOE (USA).

%

\end{document}

%% file: author-conf2005.tex
%%% Paper:    
%%% Journal:  Summer 2005 conference papers
%%% Contacts: 
%%% Non-responding authors or those who said NO are commented out.
%%% ====================================================================
%%% Click the RELOAD button on your web browser to see the updated file.
%%% ====================================================================
%%% Use \input{author} to insert this material into your latex file.
%%%%% Force institutions to appear in alphabetical order when typeset.
\affiliation{Aomori University, Aomori}
\affiliation{Budker Institute of Nuclear Physics, Novosibirsk}
\affiliation{Chiba University, Chiba}
\affiliation{Chonnam National University, Kwangju}
\affiliation{University of Cincinnati, Cincinnati, Ohio 45221}
\affiliation{University of Frankfurt, Frankfurt}
\affiliation{Gyeongsang National University, Chinju}
\affiliation{University of Hawaii, Honolulu, Hawaii 96822}
\affiliation{High Energy Accelerator Research Organization (KEK), Tsukuba}
\affiliation{Hiroshima Institute of Technology, Hiroshima}
\affiliation{Institute of High Energy Physics, Chinese Academy of Sciences, Beijing}
\affiliation{Institute of High Energy Physics, Vienna}
\affiliation{Institute for Theoretical and Experimental Physics, Moscow}
\affiliation{J. Stefan Institute, Ljubljana}
\affiliation{Kanagawa University, Yokohama}
\affiliation{Korea University, Seoul}
\affiliation{Kyoto University, Kyoto}
\affiliation{Kyungpook National University, Taegu}
\affiliation{Swiss Federal Institute of Technology of Lausanne, EPFL, Lausanne}
\affiliation{University of Ljubljana, Ljubljana}
\affiliation{University of Maribor, Maribor}
\affiliation{University of Melbourne, Victoria}
\affiliation{Nagoya University, Nagoya}
\affiliation{Nara Women's University, Nara}
\affiliation{National Central University, Chung-li}
\affiliation{National Kaohsiung Normal University, Kaohsiung}
\affiliation{National United University, Miao Li}
\affiliation{Department of Physics, National Taiwan University, Taipei}
\affiliation{H. Niewodniczanski Institute of Nuclear Physics, Krakow}
\affiliation{Nippon Dental University, Niigata}
\affiliation{Niigata University, Niigata}
\affiliation{Nova Gorica Polytechnic, Nova Gorica}
\affiliation{Osaka City University, Osaka}
\affiliation{Osaka University, Osaka}
\affiliation{Panjab University, Chandigarh}
\affiliation{Peking University, Beijing}
\affiliation{Princeton University, Princeton, New Jersey 08544}
\affiliation{RIKEN BNL Research Center, Upton, New York 11973}
\affiliation{Saga University, Saga}
\affiliation{University of Science and Technology of China, Hefei}
\affiliation{Seoul National University, Seoul}
\affiliation{Shinshu University, Nagano}
\affiliation{Sungkyunkwan University, Suwon}
\affiliation{University of Sydney, Sydney NSW}
\affiliation{Tata Institute of Fundamental Research, Bombay}
\affiliation{Toho University, Funabashi}
\affiliation{Tohoku Gakuin University, Tagajo}
\affiliation{Tohoku University, Sendai}
\affiliation{Department of Physics, University of Tokyo, Tokyo}
\affiliation{Tokyo Institute of Technology, Tokyo}
\affiliation{Tokyo Metropolitan University, Tokyo}
\affiliation{Tokyo University of Agriculture and Technology, Tokyo}
\affiliation{Toyama National College of Maritime Technology, Toyama}
\affiliation{University of Tsukuba, Tsukuba}
\affiliation{Utkal University, Bhubaneswer}
\affiliation{Virginia Polytechnic Institute and State University, Blacksburg, Virginia 24061}
\affiliation{Yonsei University, Seoul}
  \author{K.~Abe}\affiliation{High Energy Accelerator Research Organization (KEK), Tsukuba} % KEK
  \author{K.~Abe}\affiliation{Tohoku Gakuin University, Tagajo} % TohokuGakuin
  \author{I.~Adachi}\affiliation{High Energy Accelerator Research Organization (KEK), Tsukuba} % KEK
  \author{H.~Aihara}\affiliation{Department of Physics, University of Tokyo, Tokyo} % Tokyo
  \author{K.~Aoki}\affiliation{Nagoya University, Nagoya} % Nagoya
  \author{K.~Arinstein}\affiliation{Budker Institute of Nuclear Physics, Novosibirsk} % BINP
  \author{Y.~Asano}\affiliation{University of Tsukuba, Tsukuba} % Tsukuba
  \author{T.~Aso}\affiliation{Toyama National College of Maritime Technology, Toyama} % Toyama
  \author{V.~Aulchenko}\affiliation{Budker Institute of Nuclear Physics, Novosibirsk} % BINP
  \author{T.~Aushev}\affiliation{Institute for Theoretical and Experimental Physics, Moscow} % ITEP
% ...
% Lausanne
\collaboration{The Belle Collaboration}

%% file: etah_summer_2005.bbl
\begin{thebibliography}{99}
\bibitem{CC}
Throughout this paper, 
the inclusion of the charge conjugate mode decay is implied
unless otherwise stated.
\end{thebibliography}
